\documentclass[12pt]{iopart}

\usepackage[utf8]{inputenc}
\usepackage[english]{babel}

\usepackage{caption}
\usepackage{subcaption}
\usepackage{float}
\usepackage{siunitx}
\usepackage{graphicx}
\usepackage{color}
\usepackage{xcolor}
\usepackage[colorlinks=true,citecolor=blue,linkcolor=blue,urlcolor=blue]{hyperref}
\usepackage{times}
\usepackage{verbatim}
\usepackage{bm}
\usepackage{braket}
\usepackage{ragged2e}

\newcommand\RWA{\stackrel{\text{RWA}}{=}}

\begin{document}
\raggedbottom

\title{Quantum computation over the vibrational modes of a single trapped ion}

\author{Alexandre C. Ricardo$^1$ $^*$, Gubio G. de Lima$^1$ $^*$, Amanda G. Valério$^1$, Tiago de S. Farias$^1$, and Celso J. Villas-Boas$^1$}
\address{$^1$Departamento de F\'{i}sica, Universidade Federal de S\~{a}o Carlos, 13565-905 S\~{a}o Carlos, S\~{a}o Paulo, Brazil}
\address{$^*$Both authors contributed equally to this work.}
\ead{alexandre.ricardo@df.ufscar.br}





\begin{abstract} 
Continuous-variable quantum computing utilizes continuous parameters of a quantum system to encode information, promising efficient solutions to complex problems. Trapped-ion systems provide a robust platform with long coherence times and precise qubit control, enabling the manipulation of quantum information through its motional and electronic degrees of freedom. In this work, quantum operations that can be generated in trapped-ion systems are employed to investigate applications aimed at state preparation in continuous-variable quantum systems.
\end{abstract}

\maketitle

\section{Introduction}

Quantum computing in continuous-variables presents an alternative approach to the conventional qubit-based (digital) quantum computing, where qubits are enumerable and finite units of information. In the continuous model, information can be represented by conjugated variables such as position and momentum, or amplitude and phase \cite{Lloyd2003}. One of the advantages of quantum computing in this model is its ability to handle and process an infinite number of quantum states, which can make certain tasks, such as simulating complex quantum systems, to be potentially more efficient than their digital counterparts. Using this model of quantum computing in the phase-space representation of quantum mechanics, it is possible to encode the aspects of a classical neural network, as shown in \cite{PhysRevResearch.1.033063}.



Neural networks and machine learning have attracted significant attention across various scientific fields due to their versatility of applications. In physics and mathematics, these methods have found applications in a range of problems, including particle physics \cite{Kolanoski_bygubio}, high-energy physics \cite{CNN_for_highenergy_bygubio}, solving differential equations \cite{pinn_for_heatproblem_bygubio,pinn_fluid__bygubio}, and other areas \cite{938482_bygubio,DUCH199491_bygubio,9743327_bygubio}. Thus, the progress of these two promising areas, \textit{i.e.}, quantum computing and artificial intelligence, naturally led to their joining, and nowadays we can see that they have already intersected in several ways, including error mitigation in quantum circuits \cite{Bennewitz_2022, 9226505}, quantum control \cite{pinn_quantumcontrol_bygubio, Ostaszewski2019}, and quantum machine learning \cite{qnn_bygubio,schuld2015introduction_bygubio}.

One of the main efforts to adapt machine learning algorithms for quantum computers involves the exploration of Quantum Neural Networks (QNNs). In recent years, significant progress has been made to define the features and capabilities of QNNs \cite{zhao_bygubio, kwak_bygubio, zhou_bygubio}.  One possible approach is the utilization of Continuous-Variables Quantum Computing (CVQC) \cite{PhysRevLett.82.1784} to encode these models. This allows the accommodation of continuous variables of the neural network, thereby eliminating the requirement for a significant number of qubits to perform routines such as data encoding. Furthermore, the CVQC model demonstrates proficiency in applying non-linear transformations in the phase-space through the application of non-Gaussian operations, a challenging aspect for qubit-based models, which often resort to measurement post-selection \cite{barratt_bygubio} or repeated measurements in ancillary modes \cite{Bangar_bygubio,marshall_bygubio}.
In order to utilize the CVQC framework, it is necessary to employ quantum computers that are capable of encoding states in continuous variables. Photonic chips play a promising role in this context, although they exhibit limitations in the context of non-Gaussian transformations \cite{Bourassa2021blueprintscalable}. In contrast, alternative systems such as trapped ion systems have the potential to implement non-Gaussian operations \cite{Stobi_ska_2011}.

In this work, operations required for implementing continuous-variable model of quantum computing, are presented in the context of a single trapped ion. The focus of this investigation is directed towards the analysis of the behavior of the motional modes of a single ion in a Paul trap \cite{PhysRevLett.74.4091,RevModPhys.75.281} interacting with light. Using this approach, the aim is to integrate and execute quantum algorithms in CVQC, in particular it was possible to simulate a neural network model, using these techniques as a regression model and initial state preparation, to be able to simulate the entire quantum system. By focusing the analysis on the ion motion, this work is expected to provide a novel perspective on the potential of employing quantum systems for advanced computational tasks, as demonstrated by the results of the presented simulations.

\section{Continuous-variables quantum computing}\label{Sec:CVQC}

Continuous-variables quantum computing is a model which uses quantum systems described by infinite-dimensional Hilbert spaces and may be represented by continuous variables, e.g., position and momentum. This model is fundamentally different from the traditional quantum computing models described by discrete-variable systems, the qubits. In this setting, quantum information is encoded using continuous degrees of freedom of a physical system, for instance: field quadrature of light, or continuous variables that accompany the motion of trapped ions. A simple way of understanding this model is to analyze the evolution of an observable $A$ in a quantum system:
\begin{equation}\label{Obs_map}
    A(t) = e^{i\, H t}\, A(0)\, e^{-i\, Ht},
\end{equation}
\noindent for some time-independent Hamiltonian $H$. The generation of different maps in Eq.~\ref{Obs_map} varies with the different possible Hamiltonians, that represent the interactions in the system. In this context, a universal set in this model is considered to be the set of operations capable of reproducing all possible maps of $A$.

Formally, CVQC consists on the manipulation of quantum states represented by infinite-dimensional Hilbert spaces rather than finite-dimensional vectors. The manipulation of the quantum state is primarily achieved through the use of continuous-variable quantum gates, which consist of various operations known in quantum optics such as squeezing, displacement, Kerr, and phase shifts to quantum states. Therefore, quantum gates and measurements allow the execution of quantum algorithms and computations in a manner analogous to that observed in digital quantum computing.

In \cite{Lloyd2003}, it is demonstrated that a combination of linear and second-order operations in the quadrature operators $\hat{X}$ and $\hat{P}$, along with third-order or higher operations in the quadrature operators, is capable of generating any Hamiltonian expressed by polynomials in the quadrature operators. In a subsequent work by \cite{PhysRevA.99.022341}, an exact decomposition utilizing the evolution operators of $\hat{X},\,\hat{X}^2,\,\hat{X}^3$, among other operators, is presented, showcasing its practical significance due to a large reduction in the number of required operations.

\section{Continuous-variables quantum computing in a single trapped-ion} \label{Gates_sec}

This section provides an overview of the dynamics of a single trapped ion with mass $m$ in a Paul trap \cite{RevModPhys.75.281}. Using the appropriate choice of geometry, it is possible to model the light-matter interaction by assuming that the free Hamiltonian is composed of an harmonic term, characterized by the frequency $\nu$, which represents the confinement of the ion along a specific direction and a term representing the internal state of the ion, with transition frequency between the ground and excited states ($\ket{g}$ and $\ket{e}$, respectively), denoted as $\omega_0$:
\begin{equation*}
\hat{H}_0 = \hbar \nu\, \hat{a}^\dag\, \hat{a} + \frac{\hbar \omega_0}{2}\hat{\sigma}_z,
\end{equation*}
\noindent where $\hat{a}$ ($\hat{a}^\dagger$) is the annihilation (creation) operator of the motional mode on the confining direction and $\hat{\sigma}_z$ is the Pauli $Z$ operator acting on the electronic state of the ion. Thus, a dipole interaction between the system and an electromagnetic field of frequency $\omega$, wavenumber $k$ and phase $\phi$ can be represented by the interaction potential
\begin{equation*}
    \hat{V}_I(t) = \frac{\hbar}{2}\Omega\hat{\sigma}_x\exp{i\left[\eta(\hat{a}+\hat{a}^\dag)-\omega t +\phi\right]}+h.c.
\end{equation*}
The Lamb-Dicke parameter, $\eta=k\cos \theta \, \sqrt{\frac{\hbar}{2m\nu}}$, is employed in order to estimate the coupling strength between the motional and electronic states. This parameter is dependent on the angle $\theta$ between the wave vector \textbf{k} of electromagnetic field and the vibrational direction of the ion. In the interaction picture, after applying the rotating wave approximation, it is possible to obtain the interaction Hamiltonian
\begin{equation*}\label{interaction_1mode}    
    \hat{H}(t) = e^{i\hat{H}_0 t/\hbar} \, \hat{V}_I e^{-i\hat{H}_0 t/\hbar}\, \small{\RWA}  \frac{\hbar\Omega}{2}\left(\hat{\sigma}_+\, e^{-i(\delta t - \phi)}e^{i\hat{\gamma}} + \hat{\sigma}_-\, e^{i(\delta t - \phi)}e^{-i\hat{\gamma}^\dag}\right),
\end{equation*}
\noindent where $\Omega$ is the Rabi frequency, $\delta = \omega-\omega_0$ is the detuning between the laser frequency and the electronic frequency transition of the ion, and $\hat{\gamma} = \eta\left(\hat{a}\,e^{-i\nu t} + \hat{a}^\dag \, e^{i\nu t}\right)$.
 
 \subsection{Single mode operations}\label{Sec:Single_mode}

The behavior of  the motional modes of trapped ions is of great importance for quantum control and quantum computing, being an extensively investigated area \cite{PhysRevLett.76.1796,PhysRevLett.82.1971, Milburn2000801, PhysRevA.62.022311,PhysRevLett.82.1835,Lo2015}. In this section, the generation of Gaussian operations via the light-ion interaction with appropriate detunings is reviewed, following \cite{ORTIZGUTIERREZ2017166}. In the following, it is assumed that the ion's electronic state is initially prepared in the $\ket{+}$ eigenstate with eigenvalue $+1$. An alternative approach to generating interactions within motional modes, based on the parametric modulation of the harmonic trap potential, has been applied in \cite{burdsqueezing}. 

When a single ion confined in one direction interacts with a laser of frequency $\omega = \omega_0$ and null phase $\phi=0$, the effective (carrier) interaction can be written, in the Lamb-Dicke regime ($\eta\ll 1$) \cite{RevModPhys.75.281, ORTIZGUTIERREZ2017166}, as
\begin{equation}
    \label{carrier}
    \hat{H} = \frac{\hbar \Omega}{2}\hat{\sigma}_x\left(1 - \frac{\eta^2}{2}\right) - \frac{\hbar \Omega\eta^2}{2}\hat{\sigma}_x\hat{a}^\dag \hat{a},
\end{equation}
\noindent which can be seen as a phase-gate operation over the motional modes if the electronic state of the ion is an eigenstate of $\hat{\sigma}_x$. Since it was assumed that the electronic state of the ion was prepared in the $\ket{+}$ state, the first therm will represent a global phase. 

On the other hand, when a single trapped ion interacts with a bichromatic laser of frequencies $\omega = \omega_0 \pm \nu$ and non null phase $\phi$, the effective interaction can be read as
\begin{equation}
\label{displacement}
    \hat{H} = i\frac{\hbar\Omega \eta}{2} \hat{\sigma}_\phi \left(e^{i\phi/2} \hat{a}-e^{-i\phi/2} \hat{a}^\dag\right),    
\end{equation}
with $\hat{\sigma}_\phi = \hat{\sigma}_x \cos \phi + \hat{\sigma}_y \sin \phi$. This interaction can be seen as a displacement operation over the motional modes. If the ion is prepared in an eigenstate of the operator $\hat{\sigma}_\phi$, the electronic state remains uncoupled to the motional state and, thus, an arbitrary displacement can be achieved.

Finally, when the ion interacts with bichromatic light of frequencies $\omega = \omega_0 \pm 2\nu$, the effective interaction can be expressed as
\begin{equation}
    \label{squeeze}
    \hat{H} = -\frac{\hbar\Omega \eta^2}{4} \hat{\sigma}_\phi \left(e^{i\phi/2}\, \hat{a}^2 - e^{-i\phi/2}\, \hat{a}^{\dag 2}\right),
\end{equation}

\noindent which represents the squeezing operator. These operations are sufficient to perform an affine transformation in the phase space \cite{PhysRevResearch.1.033063} and also to generate Gaussian quantum states \cite{doi:10.1142/S1230161214400010,Arvind1995}, when considering a single mode. This set of operations is the first step to build a universal quantum computer that operates over continuous variables, which would require a non-Gaussian operation and a two-mode operation when considering universality in multi-mode systems.

 \subsection{Two-mode operations}
Scaling a quantum system to use more modes is necessary for larger problems, such as solving a high-dimensional partial differential equation \cite{Arrazola_2019,PhysRevA.106.052431} or building a neural network that accepts more inputs or outputs \cite{PhysRevResearch.1.033063}. Therefore, when considering an ion trapped in two (or three) directions, the interaction Hamiltonian in Eq.~\ref{interaction_1mode},  a beam splitter operation \cite{scully_zubairy_1997, mandel_wolf_1995}, united with the aforementioned operations, and a non-Gaussian interaction is sufficient to encode the aspects of universality. In the two-mode case, Eq.~\ref{interaction_1mode} is still valid with the modified operator $\hat{\gamma}$:
\begin{equation}
    \hat{\gamma} = \eta_x(\hat{a}\, e^{-i\nu_x t}+ \hat{a}^\dag \, e^{i\nu_x t}) + \eta_y(\hat{b}\, e^{-i\nu_y t}+ \hat{b}^\dag \, e^{i\nu_y t}),
\end{equation}

\noindent where $\nu_x\mbox{ } (\nu_y)$, $ \hat{a}\mbox{ } (\hat{b}) $ and  $ \hat{a}^\dagger\mbox{ } (\hat{b}^\dagger) $  are the frequency, annihilation and creation operators associated to the $x\mbox{ } (y)$ -mode, respectively. Setting a bichromatic laser with frequencies $\omega = \omega_0 \pm(\nu_x - \nu_y)$, the effective Hamiltonian can be written as 
\begin{equation}
    \label{BS}
    \hat{H} = -\frac{\hbar \Omega \eta_x\eta_y}{2}\left(\hat{a}\,\hat{b}^\dag + \hat{b}^\dag \hat{a}\right)
\end{equation}
\noindent and the corresponding evolution operator can be regarded as the beam splitter operation with transmission parameter $\theta(t) = \Omega \eta_x\eta_y t/2$. Another interesting operation with applications to quantum sensing \cite{PhysRevA.103.062405} and generation of highly entangled Gaussian states \cite{Ping-Yun-Xia_2008} is the two mode squeezing, generated by choosing $\delta = \pm(\nu_x+\nu_y)$. For this choice, the Hamiltonian is expressed as follows:
\begin{equation}\label{TMS}
    \hat{H} = -\frac{\hbar \Omega \eta_x \eta_y}{2}\hat{\sigma}_\phi\left(\hat{a}\hat{b}\, e^{i\phi}+\hat{a}^\dag\hat{b}^\dag\, e^{-i\phi}\right).
\end{equation}

The set of operations comprised in Sec.~\ref{Sec:Single_mode} together with any of the operations in this section are sufficient to prepare any Gaussian state in a system formed by different quantum modes.

\subsection{Non-Gaussian operations}

To achieve universality — defined as the capability to approximate the dynamics of any analytical Hamiltonian $\hat{H}$ — it is appropriate to employ a combination of the Gaussian operations together with a non-Gaussian operation \cite{Lloyd2003,PhysRevA.99.022341}. The objective of this section is to provide a comprehensive overview of non-Gaussian interactions that can be generated in a trapped-ion system.

The first interaction discussed in this section is the trisqueezing \cite{Eriksson2024, băzăvan2024squeezing}, which naturally emerges as a potential candidate for non-Gaussian operation. It is generated by the interaction between the system and a bichromatic field with $\delta = \pm 3\nu$. In this case, it is possible to achieve the effective interaction
\begin{equation}\label{trisqueeze}
    \hat{H} = \frac{\hbar \Omega \eta^3}{12}\hat{\sigma}_x \left(\hat{a}^{\dag 3}\, e^{i\phi} + \hat{a}^3\,e^{-i\phi}\right).
\end{equation}
This non-Gaussian interaction produces and annihilates a high number of phonons in the vibrational modes, generating a non-linear effect in the phase-space representation. Furthermore, this interaction can be generated through the use of spin-dependent forces in a similar way \cite{PhysRevA.104.032609}.

Considering the carrier interaction ($\delta=0$), the fourth-order expansion of Eq.~\ref{interaction_1mode} results in
\begin{equation}\label{Kerr}
    \hat{H} = -\frac{\hbar\Omega}{2}\left(\eta^2-\frac{\eta^4}{2}\right)\hat{\sigma}_x\hat{a}^\dag\,\hat{a}+\frac{\hbar \Omega \eta^4}{8}\hat{\sigma}_x \hat{a}^{\dagger 2} \hat{a}^2,
\end{equation}
\noindent which can be regarded as a Kerr interaction, 
\begin{equation}
    \hat{H} = \frac{\hbar \Omega \eta^4}{8}\hat{\sigma}_x \hat{a}^{\dagger 2} \hat{a}^2,
\end{equation}
\noindent in a rotating frame with the correct frequency \cite{Stobi_ska_2011}. In the following applications for quantum computing, the effective Rabi frequency will be corrected in second order to
\begin{equation}
    \Omega \approx \Omega_0 e^{-\eta^2},
\end{equation}
\noindent according to \cite{RevModPhys.75.281}. This correction is particularly important when considering higher Lamb-Dicke parameters to obtain a higher fidelity in the dynamics, allowing the simulations to achieve better results in this operation. 

\section{Applications}\label{Sec:Applications}

This section presents the results of numerical simulations of the operations discussed applied in different contexts of quantum algorithms. Firstly, a numerical simulation was conducted on the continuous variable system to evaluate the fidelity of the quantum states produced by the operations delineated in Sec.~\ref{Gates_sec}. Potential uses of such states include, but are not limited to, quantum error correction \cite{PhysRevA.64.012310} and quantum sensing \cite{Guo2020,PhysRevA.103.062405}. As a second application, we use the quantum neural network presented in \cite{PhysRevResearch.1.033063} with the same operations for a regression problem. 
As a final application, the same quantum neural network model is developed using the derived interactions and applied to the state preparation problem.

\begin{figure}[!ht]
\begin{subcaptiongroup}
\includegraphics[clip, width=\columnwidth]{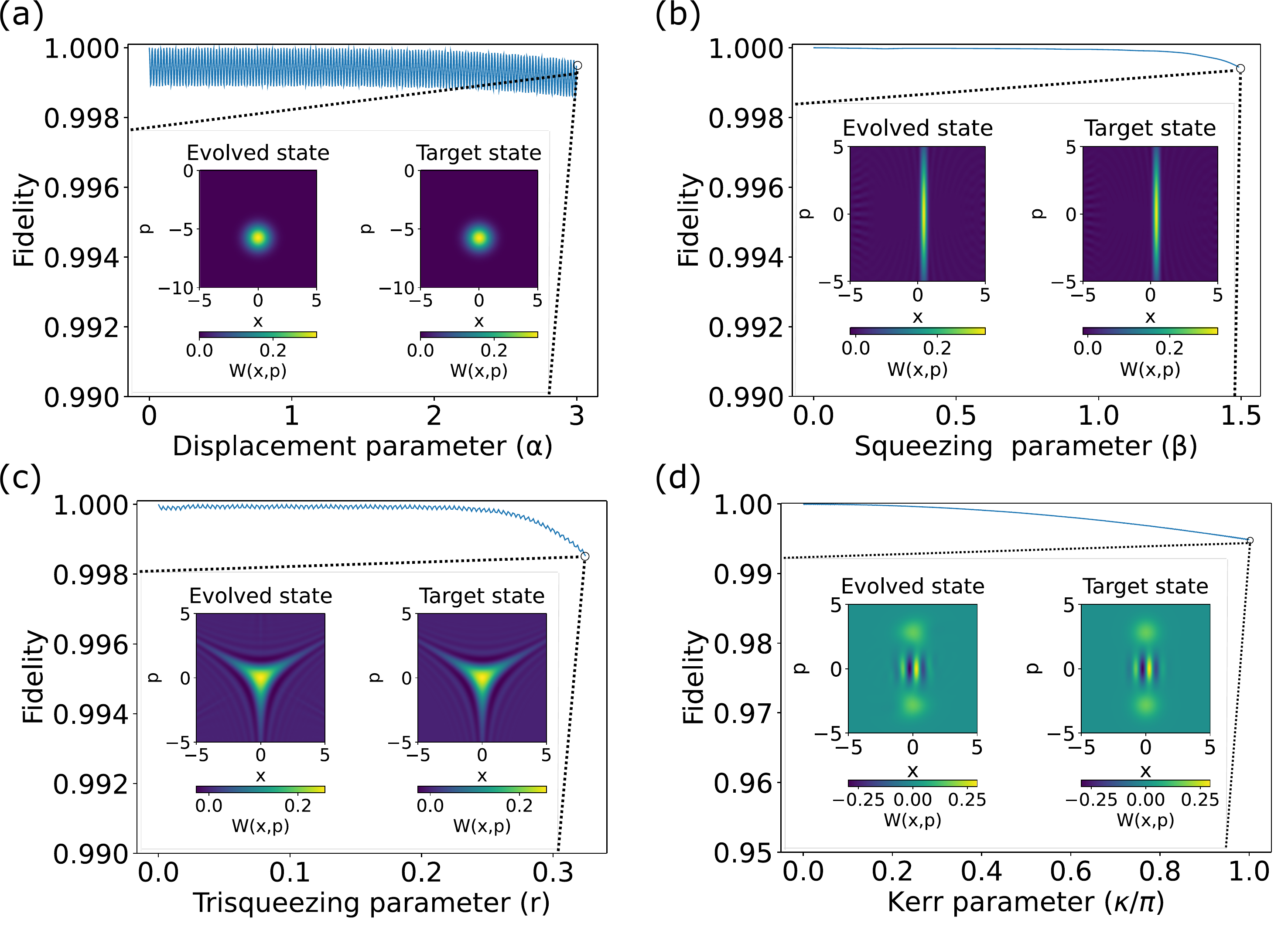}
    \phantomcaption \label{subfig-displacement}
    \phantomcaption \label{subfig-squeeze}
    \phantomcaption \label{subfig-trisqueeze}
    \phantomcaption \label{subfig-kerr}
\end{subcaptiongroup}
    \caption{\justifying Fidelity, defined as the module squared of the overlap between two quantum states for different states as a function of the parameters used in quantum operations. Each fidelity plot includes an inset displaying the Wigner quasi-probability distributions $W(x,p)$, where the horizontal (vertical) axis represents the quadratures of position (momentum).  The target $W(x,p)$ can be seen on the right side of the inset, while the left side shows the $W(x,p)$ after the evolution of the full Hamiltonian, derived from trapped ion dynamics. For every operation, the values of trap frequency and Rabi frequency were defined to be $\nu = 2\pi\times 3$ MHz and $\Omega = 2\pi \times 100$ kHz, respectively. In each panel, the fidelity as a function of the value of the operation parameter is presented. \subref{subfig-displacement} Coherent states with final fidelity equal to $ 0.9994$, for $\alpha = 3.0$; \subref{subfig-squeeze} Squeezed states with final fidelity equal to $0.9994$ for $\beta= 1.4$. \subref{subfig-trisqueeze} Trisqueezed states with final fidelity equal to $0.9985$ for $r= 0.32$. \subref{subfig-kerr} Cat states with final fidelity equal to $ 0.994$ for $\kappa/\pi= 1$.}
    \label{Fig:Fid}
\end{figure}

 \subsection{Generation of quantum states}

The interactions presented in the previous section were employed here to generate specific quantum states, with the objective of estimating their fidelities and facilitating the validation of the theoretical model. We start with quantum coherent states, that are a fundamental concept in quantum mechanics that exhibit properties similar to classical states, characterized by well-defined phase and minimum uncertainty in position and momentum. Mathematically, coherent states are eigenstates of the annihilation operator $\hat{a}$, making them significant in the context of continuous quantum information. Quantum coherent states have been employed in various fields and are easily generated in the context of several different quantum experiments.
In Fig.~\ref{Fig:Fid}, panel \subref{subfig-displacement} displays the fidelity of the coherent states generated by applying the displacement operator described by Eq.~\ref{displacement}.

Squeezed coherent states represent a class of non-classical states that exhibit reduced uncertainty, below the Heisenberg limit, in one observable at the expense of increased uncertainty in another. These states emerge as a result of applying a squeeze operator on coherent states, effectively altering their uncertainty properties. Squeezed coherent states have attracted considerable interest because of their potential applications in quantum-enhanced measurement schemes, such as gravitational wave detection \cite{DIMOPOULOS200937} and high-precision interferometry \cite{Caves:86}. Panel~\subref{subfig-squeeze} of Fig.~\ref{Fig:Fid} depicts the fidelity of squeezed coherent states generated by applying the squeezing operation in a coherent state to assess its fidelity. Panel~\subref{subfig-trisqueeze} in Fig.~\ref{Fig:Fid} refers to the fidelity of creating trisqueezed states \cite{PhysRevA.35.1659} to estimate the fidelity of the trisqueezing operation. In the same way, panel~\subref{subfig-kerr} in Fig.~\ref{Fig:Fid} displays the fidelity of the state generated through the Kerr operation applied on an initial coherent state. In particular, the last step in the time evolution refers to the creation of a cat state.

It is also important to pay attention to the number of excitations in the vibrational modes since the manipulation of states with a high expectation value of the number operator $\hat{n}$ using these operations with high fidelity is significantly harder since the Lamb-Dicke regime is no longer satisfied and the approximations made in Sec.~\ref{Gates_sec} do not hold. For this, and due to the high cost of simulating systems with a high number of excitations, the applications in this work consider a small number of gates that will cause a small variation in the number of excitations in the mode.

\subsection{Quantum neural network model}

Previously, the generation of states from single gates has been employed to produce specific quantum states, each requiring a single application of specific operations to verify their fidelity. Nevertheless, the generation of a general quantum state in the Hilbert space remains a significant challenge. A promising way to accomplish such a task is to integrate quantum machine learning techniques, particularly through the utilization of parameterized quantum circuits. Hence, a quantum neural network model \cite{Arrazola_2019} was implemented, considering the interactions generated in this paper and applying the Kerr interaction Eq.~\ref{Kerr} as an activation function in the quantum circuit.

The decision to use a Kerr interaction instead of the trisqueezing gate described in Eq.~\ref{trisqueeze} is based on both numerical and experimental considerations. Firstly, the trisqueezing interaction generally leads to an increase in the average phonon number in the quantum mode. This has two implications: it makes the classical simulation of the quantum system more challenging, and it reduces the precision of the rotating-wave approximations. Secondly, as the average phonon number increases, the measurement of the states becomes more difficult, even for state-of-the-art methods \cite{PhysRevLett.131.223603}. 

\begin{figure}[H]
    \begin{subcaptiongroup}
    \centering
    \includegraphics[trim={.25cm 0 2cm 0},clip, width=.49\columnwidth]{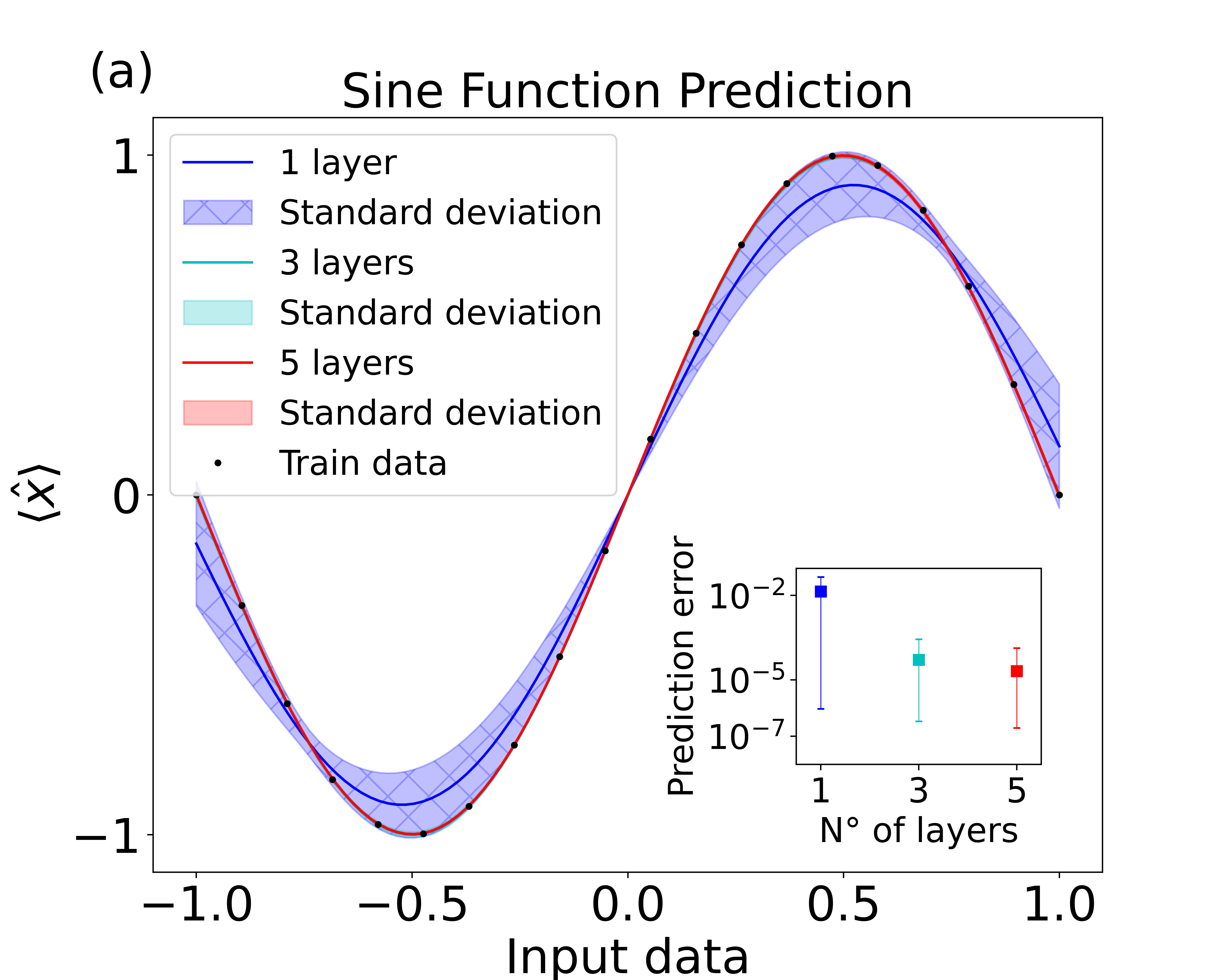}
    \includegraphics[trim={.25cm 0 2cm 0},clip, width=.49\columnwidth]{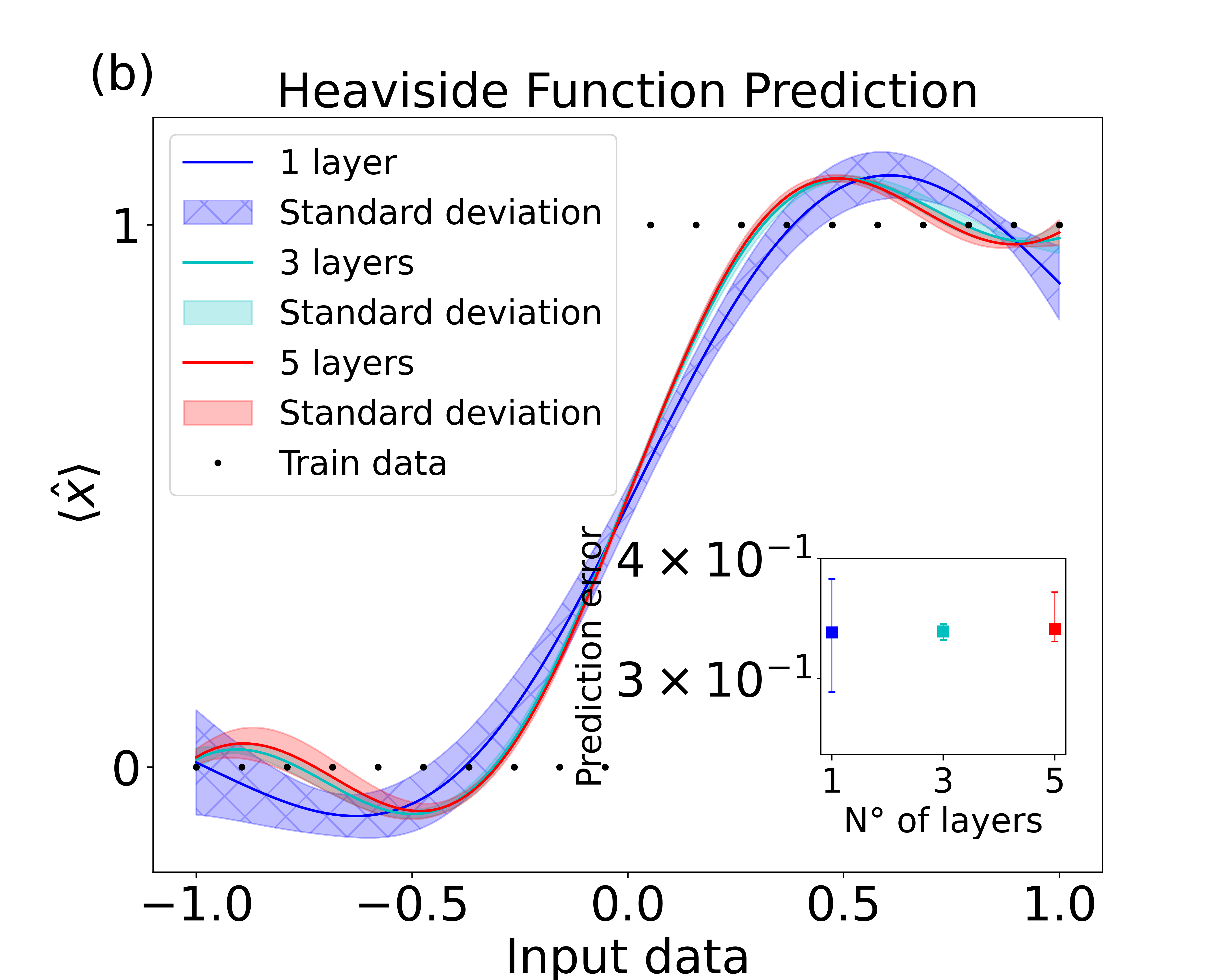}
    \phantomcaption \label{subfig:sine}
    \phantomcaption \label{subfig:heaviside}
    \end{subcaptiongroup}
    \caption{\justifying Results obtained by the regression problem using the QNN for different number of layers. The horizontal axis represents the input data used in the encoder before the QNN, while the vertical axis represents the expected value of the operator $\hat{X}$. The black dots represent the target functions, and the solid curves represent the average of the outputs of the QNN for 11 different parameter initializations. The shaded region illustrates the standard deviation, i.e. a $68\%$ confidence interval, across the different executions, reflecting the variability due to different parameter initializations. The inset in each panel presents the mean square error and its standard deviation between the test data and the outputs of the quantum model for each layer after training. In panel~\subref{subfig:sine}, the target function is the sine function $f(x)=\sin(\pi x)$ and in panel~\subref{subfig:heaviside}, the target function is the Heaviside step function.}
     \label{fig:regression}
\end{figure}

In this section, the quantum algorithm is simulated by numerically evaluating the full Hamiltonian dynamics of Eq.~\ref{interaction_1mode} for each interaction. This evaluation aims to investigate how the interactions discussed in Sec.~\ref{Gates_sec} would perform in the QNN model, beginning with a classical regression task, depicted in two different scenarios in Fig.~\ref{fig:regression}. In the first scenario, the quantum model predicts a continuous, oscillatory curve, while in the second case, the target function is set to be the discontinuous Heaviside distribution. It is observed that the oscillatory function exhibits variance getting smaller across distinct initialization when increasing the number of layers, suggesting that the model is capable of converging toward an optimal solution. However, the system cannot fit the Heaviside function within the same number of layers and optimization steps, highlighting the expected behavior of the QNN regarding discontinuous functions, even when increasing the number of layers.

\begin{figure}[H]
    \centering
    \includegraphics[clip,trim = {1.1cm, 1cm, 1.2cm, 0cm},width=\linewidth]{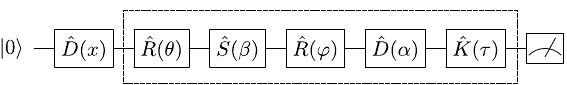}
    \caption{\justifying Sequence of operations required to execute a single-mode, single-layer quantum neural network according to the model proposed in \cite{PhysRevResearch.1.033063}. The variable $x$ encodes the input, which in the case of a regression problem, is the variable $x$ of the function $f(x)$ that the neural network tries to fit. The set of parameters $\{\theta,\,\varphi, \,\beta,\, \alpha,\,\tau\}$ assigned to phase-gate (twice), squeezing, displacement and Kerr operations, respectively, and are optimized according to the measurements of the mode quadrature $\hat{X}$. The number of layers can be increased by repeating the operations within the dotted rectangle with different parameters.}
    \label{fig:circ}
\end{figure}

In the QNN model \cite{PhysRevResearch.1.033063} applied to execute the regression task, the information is encoded in the quadrature operator $\hat{X}$. After using the displacement operation to create a feature map that encodes classical data into the quantum phase space, the information is processed through a sequence of Gaussian operations designed to perform an affine transformation that effectively acts as the neural network's weights and biases over the phase space, followed by a non-Gaussian operation representing the activation function. This variational circuit, depicted in Fig.~\ref{fig:circ},  is equivalent to a single quantum neuron in this model. Additional layers can be added by repeating the sequence of operations after the encoding displacement. The output of the network is the expected value of the $\hat{X}$ operator, which is compared to the behavior of the function that is expected to fit. 

The training process aims to achieve the optimal fit between the output given by the neural network and the target function. To accomplish this, a loss function defined by the mean square error between the outputs of the neural network and the exact data is defined and minimized using classical optimization methods. In this work, the Adaptative Moment Estimation Algorithm (Adam) \cite{kingma2017adam} is chosen to be the classical optimization subroutine. Due to the continuous and oscillatory nature of the quantum quadratures, it is expected that the model will be more adept at identifying and fitting oscillatory curves, while discontinuous functions are less likely to be fitted, resulting in effects similar to the Gibbs phenomenon \cite{Vretblad2003-oe} as the number of layers increases. 

In addition to the behavior of the function to be fit, several factors may impact the quality of the output and the difficulty of training the neural network. In particular, Fig.~\ref{fig:regression} shows the effect of different parameter initializations. It is expected that as the initial parameters are altered, the classical optimization process tends to change the direction in the parameter space, thereby modifying the optimal parameters given by the optimizer.

\subsection{State preparation}

As a second application, the focus is on quantum state preparation through the generation of random quantum states. In this section, we use the variational circuit presented in Fig.~\ref{fig:circ} to approximate quasi-random states with non-zero amplitudes of the first $N=20$ Fock states in one vibrational mode. Effectively, the goal is to obtain, after optimization, normalized states that can be written as

\begin{equation*}
    \ket{\psi} = \sum_{i=0}^{19} c_i \ket{i},
\end{equation*}

\noindent where $\ket{i}$ are Fock states and the random amplitudes $\{c_i\}$ were generated by sampling uniformly complex numbers, imposing normalization. In the process of state preparation, the model's cut-off dimension was established at $N=20$, and this is why we are refering to quasi-random states.  However, to address the challenges associated with the cut-off dimension for continuous variables, we modulated the amplitudes of the states by a random normal distribution, resulting in higher amplitudes for basis states closer to $|0\rangle$. 

To evaluate the efficacy of our quantum state preparation process, the overlap fidelity metric, defined as $F=|\langle \psi | \phi \rangle|^2$, was employed. Fidelity provides a quantitative measure of the closeness between the prepared quantum state $|\phi\rangle$, which is the output of our model, and the target random state $|\psi\rangle$. The objective of optimizing the model is maximizing this fidelity, which is indicative of the model’s accuracy in reproducing the target quantum state.

\begin{figure}[!ht]
    \begin{subcaptiongroup}
    \centering
    \includegraphics[width=.49\columnwidth]{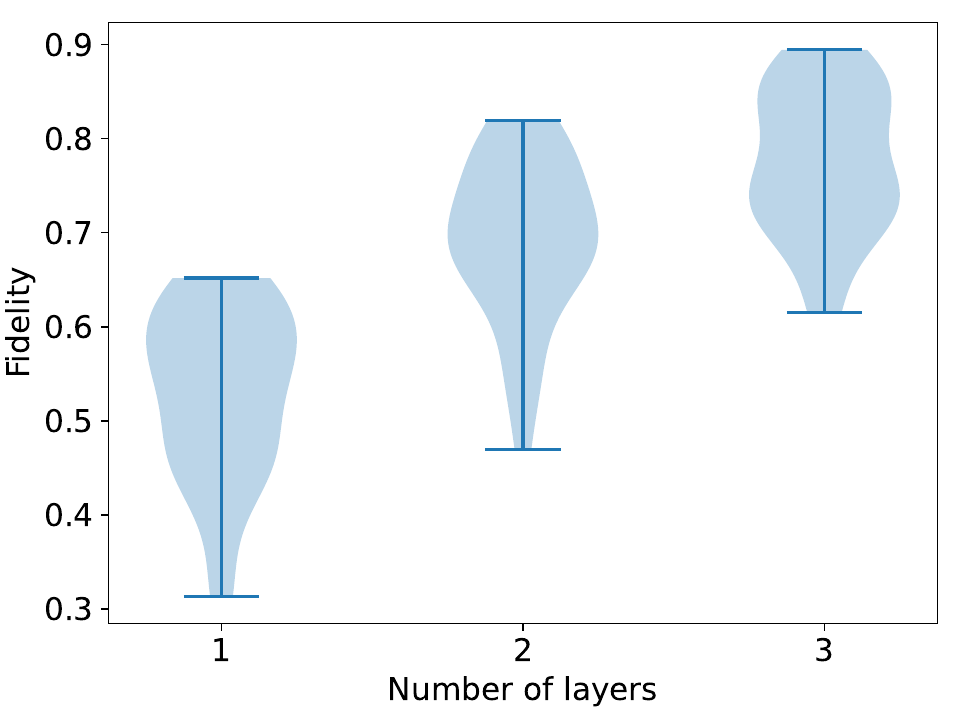}
    \includegraphics[width=.49\columnwidth]{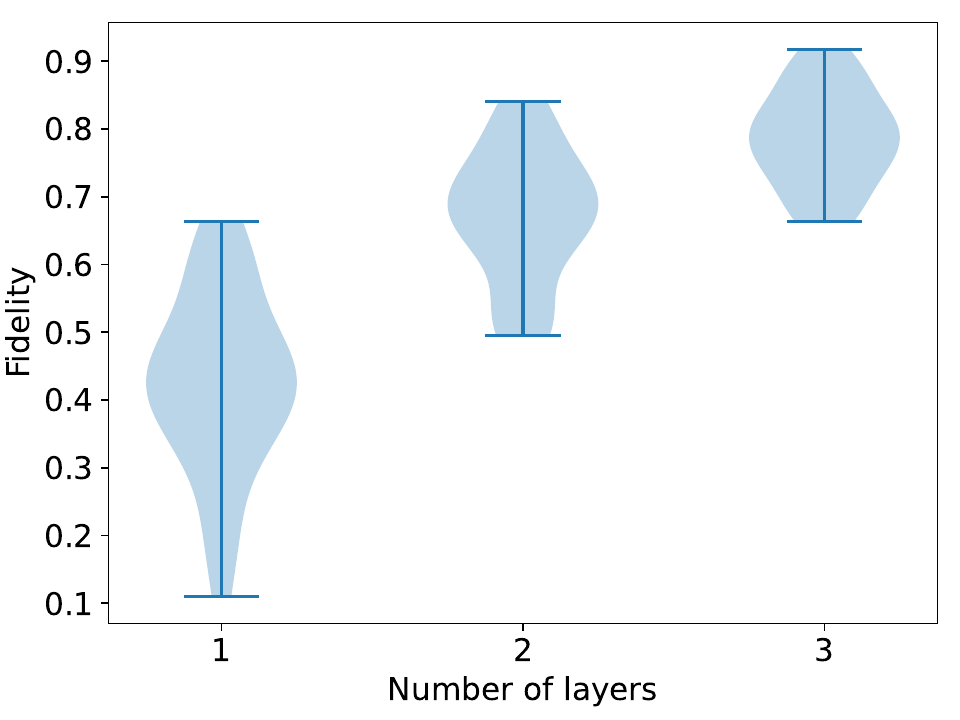}
    \phantomcaption \label{subfig:singletarget}
    \phantomcaption \label{subfig:multipletarget}
    \end{subcaptiongroup}
    \caption{\justifying  ~\subref{subfig:singletarget} Violin plots illustrating the distribution of fidelity between a single target state and the state prepared by the model across multiple parameter initializations and varying numbers of layers. The horizontal axis represents the number of layers in the quantum neural network, while the vertical axis indicates the fidelity. Each violin plot captures the fidelity distribution over 30 random initializations for models with one to three layers. An improvement in fidelity is observed with an increasing number of layers, with three-layer models achieving fidelities of up to $0.9$. \subref{subfig:multipletarget} Violin plots illustrating the fidelity distribution for preparing multiple random target states using quantum neural networks with varying numbers of layers. The horizontal axis represents the number of layers in the QNN, while the vertical axis indicates fidelity. These plots display the minimum, maximum, and overall spread of fidelity scores for each layer configuration. The results show that as the number of layers increases, the fidelity improves, highlighting the enhanced performance of deeper networks in accurately preparing random quantum states.}
    \label{fig:singletarget}
\end{figure}

The training process is driven by the quadratic difference between fidelity and its maximum achievable value, defined by the loss function $\mathcal{L}=(1-F)^2$. This loss function effectively captures the deviation from perfect fidelity, with the optimization process focused on minimizing this loss as much as possible. Through iterative training and model optimization, the goal is to develop a state preparation process capable of generating random quantum states with high fidelity.

A single random state was generated to investigate the sensitivity and capability of the model in this context. Subsequently, several models were trained, using $30$ different uniformly distributed random initialization parameters and varying the number of layers from one to three. To facilitate the optimization process across these varied models, we employed the Adam optimizer, with the hyper-parameters set to a learning rate of $\eta=0.001$, and the exponential decay rates for the first and second moment estimates configured as $\beta_1=0.9$ and $\beta_2=0.999$, respectively.
This approach permitted a systematic analysis of the effects of various initial conditions and architectural configurations on the performance of the models in preparing the target random state. Panel~\subref{subfig:singletarget} of Fig.~\ref{fig:singletarget} presents the distribution of the fidelity for the different initializations and layer configurations using violin plots \cite{violinplot}. These violin plots show the distribution of the data by depicting the probability density at different fidelity values. The width of the plot at each point reflects the density, with wider sections indicating a higher concentration of values, providing a view of the spread and concentration of fidelity values across different configurations. As the number of layers increases, the fidelity also improves, reaching up to $F=0.9$ for models with three layers.

\begin{figure}[H]
    \centering
    \begin{subfigure}{0.4\textwidth} 
        \centering
        \caption*{(a)}        
        \includegraphics[clip,trim={.5cm, 0 0 0},width=1.1\linewidth]{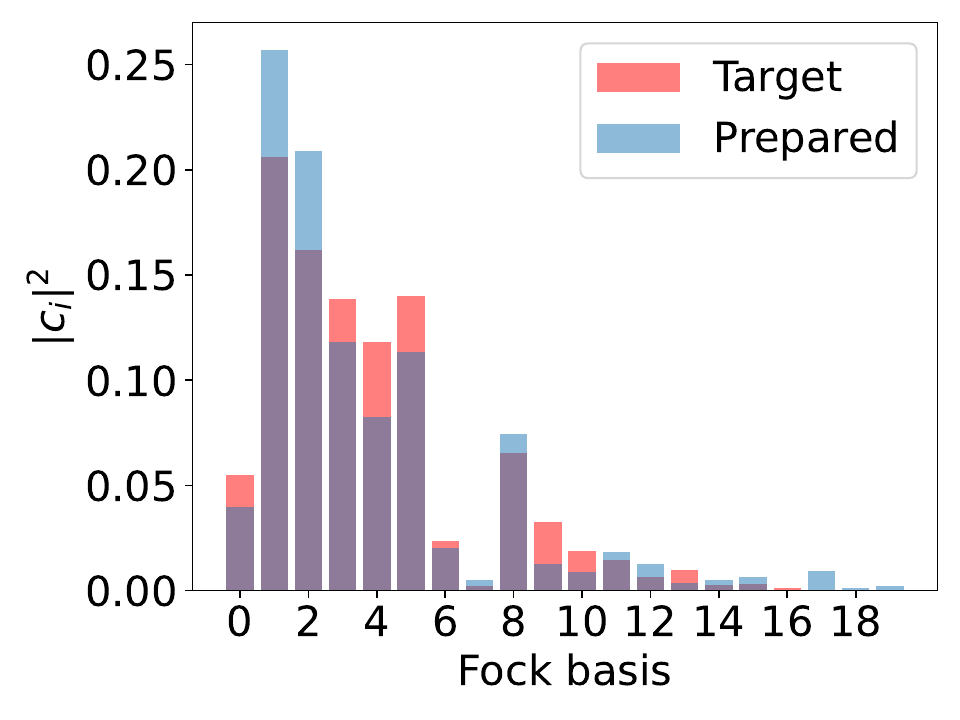}
        \phantomcaption \label{subfig:population}
    \end{subfigure}
    \begin{subfigure}{0.5\textwidth} 
        \centering
        \begin{subfigure}{1\linewidth}
            \centering
            \caption*{(b)}    
            \includegraphics[width=0.75\linewidth]{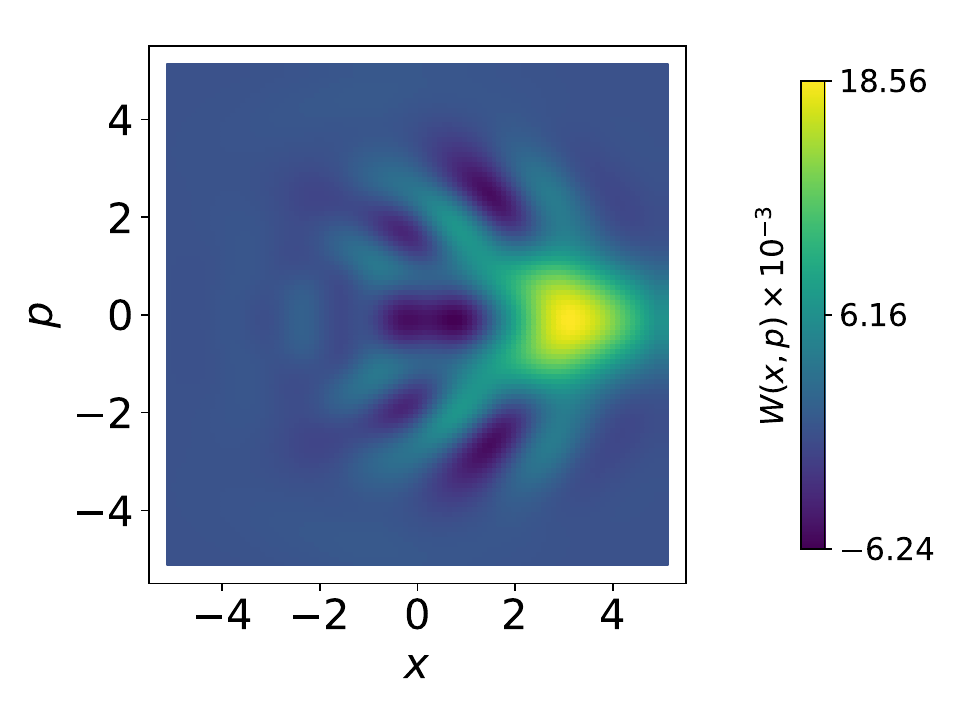}
            \phantomcaption \label{subfig:wignera}
           
        \end{subfigure}
        \begin{subfigure}{1\linewidth}
            \centering
            \caption*{(c)}        
            \includegraphics[width=0.75\linewidth]{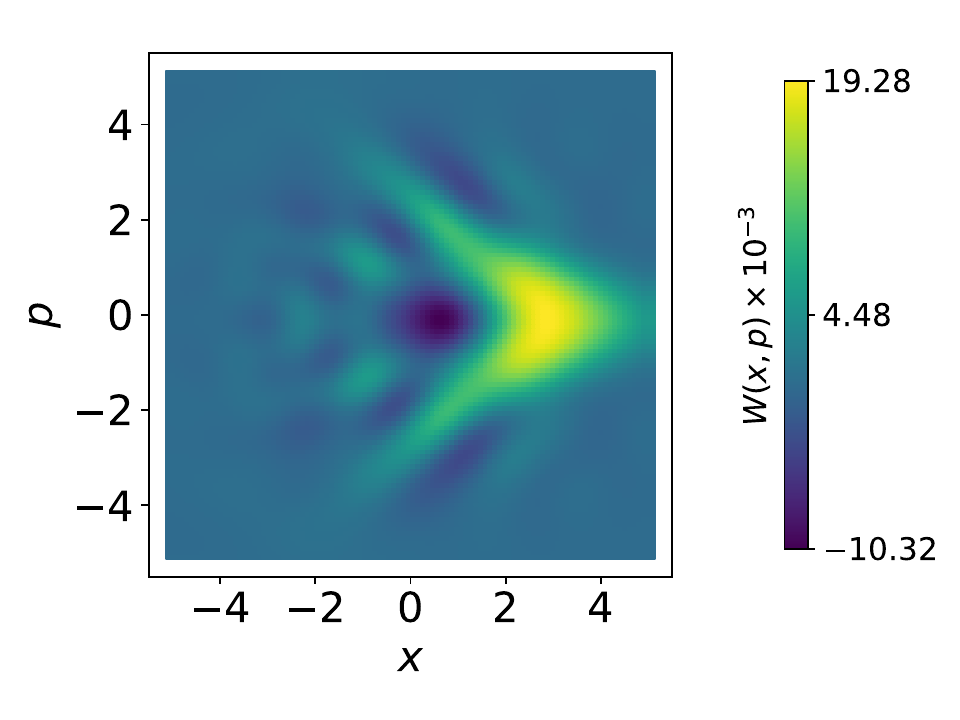}
            \phantomcaption \label{subfig:wignerb}
        \end{subfigure}
    \end{subfigure}
    
    \caption{\justifying Visualization of the process for target state preparation, where the model successfully achieved a fidelity of $0.9$. The figure is organized into three subplots: the vibrational state population on the top panel~\subref{subfig:population}, and the Wigner function distribution on the bottom, panels~\subref{subfig:wignera} and \subref{subfig:wignerb}, both plotted across their respective quadrature position and momentum axes. The upper panel compares the populations of the target state (red bars) and the state prepared by the neural network (blue bars), enabling a direct comparison between the target random state. In the bottom panels, the Wigner distributions of the target state and the state generated by the quantum neural network are displayed respectively on panels~\subref{subfig:wignera} and \subref{subfig:wignerb}.}
    \label{fig:randomstatevis}
\end{figure}

Panel~\subref{subfig:multipletarget} of Fig.~\ref{fig:singletarget}, on the other hand, illustrates the fidelity achieved in the preparation of multiple random states using quantum neural networks with varying numbers of layers. The violin plots depict the distribution of fidelity scores, showing the minimum, maximum, and overall spread of the fidelity values observed. Consistent with the results obtained for the preparation of a single random target state, an increase in the number of layers corresponds to an improvement in fidelity. Further details regarding the state preparation that achieved the highest fidelity can be found in Fig.~\ref{fig:randomstatevis}.

\section{Conclusions} \label{Conclusions}

This work outlines strategies for processing continuous-variable information through light-matter interactions in trapped-ion systems, allowing one to engineer a universal set of operations. With such a set, universal quantum computing based on continuous variables can be implemented, and one can generate and manipulate arbitrary non-Gaussian continuous-variable states using this platform. To this end, we generated various Gaussian operations with arbitrary parameters, including Displacement, Squeezing, and Phase Gates \cite{Su_2013}, as supported by prior research \cite{ORTIZGUTIERREZ2017166}. Additionally, two different non-Gaussian operations were introduced: The Trisqueezing and Kerr interactions. The Kerr interaction is based on the method displayed in a previous work \cite{Stobi_ska_2011}, incorporating a second-order correction on the Rabi frequency that improves its fidelity.

We simulated these interactions using optimal parameters, achieving fidelities exceeding $99\%$ in every case, including the non-Gaussian operations. Furthermore, we demonstrated the applicability of these simulated interactions in quantum algorithms. Specifically, we were able to model a quantum neural network that could approximate smooth data distributions as well as normalized quasi-random states, using fidelity as both the primary evaluation metric and the loss function. 

Our findings suggest that increasing the number of layers in the quantum neural network improves the fidelity of the prepared states, achieving values as high as $F = 0.9$. In essence, increasing the number of layers, thereby expanding the size of the circuit, one can significantly enhance the fidelity of state preparation.

Our results represent a significant step toward the realization of universal continuous-variable quantum computing, an important achievement explored in different platforms \cite{Pan_2023,Eriksson_2024,Takeda2019-nm}, each with its limitations. Although it is possible to confine more than one ion in the same trap, thus increasing the number of modes, scaling up the number of ions causes the frequencies of the different collective modes to become closer together. Consequently, distinguishing the individual sidebands of each mode becomes challenging, an aspect that demands deeper investigation.

\section{Acknowledgments}
This work was supported by the Coordenação de Aperfeiçoamento de Pessoal de  Nível Superior (CAPES) - Finance Code 001 and São Paulo Research Foundation (FAPESP) grants No. 2022/00209-6, 2023/14831-3 and 2023/15739-3. C.J.V.-B. is also grateful for the support by the National Council for Scientific and Technological Development (CNPq) Grants No. 465469/2014-0 and 311612/2021-0. This work is also part of the CNPq Grants No. 140467/2022-0 and 139701/2023-0.

\newpage
\section*{References}
\bibliographystyle{iopart-num}
\bibliography{refs1}

\end{document}